# An Empirical Study of Teaching Methodologies and Learning Outcomes for Online and in-class Networking Course Sections

**Dr. Pouyan Ahmadi,** *George Mason University*

Pouyan Ahmadi is an Assistant Professor in the Department of Information Sciences and Technology. His research interests include cooperative communications and networking, cross-layer design of wireless networks, relay deployment and selection in wireless networks. In 2013, Dr. Ahmadi received the Best Graduate Student Paper Award at Wireless Telecommunications Symposium (WTS). He also won the Presidential Scholarship Award in 2010 at GMU. Dr. Ahmadi earned his Ph.D. in Electrical and Computer Engineering at George Mason University, M.S. in Architecture of Computer Systems at Iran University of Science and Technology, and B.S. in Computer Engineering at Azad University.

**Dr. Khondkar Islam,** *George Mason University*

Khondkar Islam is an Associate Professor and Associate Chair for Undergraduate Studies in the Department of Information Sciences and Technology. Dr. Islam is the Coordinator for the Networking Concentration. He is also the founder and director of the Virtual Academy (VirtAc) at George Mason University (GMU). His research interests include distributed and peer-to-peer systems, overlay and wireless networks, network security, and distance education for instructor training and student learning. Dr. Islam earned his Ph.D. and B.S. at George Mason University, and M.S. at American University.

**Mr. Salman Yousaf,** *George Mason University*

Salman Yousaf is a graduate student in the field of Data Analytics Engineering. His research interests include learning analytics, big data, distributed machine learning and distance education for instructor and student learning. Salman earned his bachelor's at National University of Sciences and Technology and worked in the telecom industry in multiple roles such as Radio Optimization Engineer, Cell Planning Engineer and Technical Performance Analyst.

**Abstract**

To enhance student learning, we demonstrate an experimental study to analyze student learning outcomes in online and in-class sections of a core data communications course of the Undergraduate IT program in the Information Sciences and Technology (IST) Department at George Mason University (GMU). In this study, student performance is evaluated based on course assessments. This includes home and lab assignments, skill-based assessment, and traditional midterm exam across all 4 sections of the course. All sections have analogous content, assessment plan and teaching methodologies. Student demographics such as exam type and location preferences that may play an important role in their learning process are considered in our study. We had to collect vast amount of data from the learning management system (LMS), Blackboard (BB) Learn, in order to compare and study the results of several assessment outcomes for all students within their respective section and amongst students of other sections. We then tried to understand whether demographics have any influence on student performance by correlating individual student's survey response to his/her performance in the class. The numerical results up to mid-semester reveal remarkable insights on student success in the online and face-to-face (F2F) sections.

**Keywords**

Assessment methods, statistical analysis, demographics, distance learning

## 1. Introduction

Over the last few decades, there has been increasing awareness on emerging teaching techniques to align assessments with learning objectives. The ability of various assessment strategies to enhance student performance is influenced by various sources. Black and Wiliam [1] in their pivotal investigation use the term assessment for all activities conducted by teachers for students, in evaluating themselves served as feedback to the instructors. However, to characterize assessment, which still has a vague definition, we need more comparative data from studies.

In this study we compare 4 sections of the same course offered in the same semester. One of these sections is offered online. All 4 sections use the same learning management system (LMS) called Blackboard™ (BB) for content delivery, assignments, announcements, discussion forums, and grade book. BB is the official LMS of GMU and it is used by instructors to upload the course materials (e.g. lecture slides, lecture and lab videos for the distance learner, syllabus, etc.), conduct quizzes, and communicate with the students. Students use it to download and submit their assignments, complete the quizzes, access course materials, and check their grades.

Examinations are primarily the standard student assessment techniques and most of the undergraduate courses evaluate student learning at mid and at end of semester. Since examinations are primary assessment methods, the focus of this study is on three types of questions: multiple choice (MC), fill-in-the-blank and essay questions. Moreover, since students are continually shaped by the environment and the variables from which they enclose themselves [2], we have factored in student demographics and correlated this data with student performance.

Section 2 covers our extensive literature review on the effect of different assessment methods on student learning and performance. We elaborate on the data and methods used in our study in section 3. Data analysis with numerical results is presented in section 4. Concluding remarks and future work are discussed in section 5.

## 2. Literature Review

Demand for methodical and efficient assessment has been on the rise over the past decade. This is especially true in the higher education arena. Assessment is crucial for evaluating the quality of students' achievement [3]. Academicians suffer from ongoing doubts regarding the effectiveness of assessment, on whether assessments really apply their teaching methods, enhance student learning outcomes and/or the items on the syllabi [4]. One of the reasons why the assessment has had little influence on the teaching-learning procedure is lack of studies regarding the relation between the assessment type and student performance. Understanding this association meticulously would lead to evolving assessment approaches that could undeniably benefit student learning process.

Problem-based learning from the viewpoint of assessment is also analyzed in Gijbels et al [5]. They suggest that the associations of assessment must be taken into account in observing the effects of problem-based learning, perhaps in all equivalent educational research. In [6], an empirical study is conducted to understand student activities within the system, and conclude their relation with student performance. The frequency of accessing the material and the time duration expended on assessments are also considered in their study. They conclude that there is a link between students' use of Blackboard and student performance, but there was no evidence of noteworthy difference across a variety of demographic aspects.

Simikin et al in [7] investigate the connection between MCQs and student understanding. They propose a model for comparing MCQ tests and other type of questions where students write their own answers in essay format, rather than selecting answers from predetermined choices. They validate that properly constructed MCQ examinations may solve some of the problems that have been identified in these type of exams. Scouller [8] investigates the effect of assessment technique on student learning, where MCQ examination is compared with essay assignments. Via a survey-like approach, issues related to student's preparation and perceptions of two procedures of assessment of the same course was examined. Results show discrete patterns that employ different methods of assessments. According to this study, lack of good performance in the MCQ examination was due to utilizing profound learning strategies. On the contrary, when students prepare for their assignment essays, they expect to employ deeper learning methods which assess advanced stages of intellectual processing.

In more recent studies, educational social media is used as a tool for learning. For instance, in [9], authors managed to deliver better insight into how social collaboration and participation seemed to be an important factor in students' performance. The students were asked about their attitudes, insights, and use of the technology through an end-of-class survey. Their findings suggest that students with lower performance were less motivated to work with others, whereas students with more inclination to work with others had higher performance.

Another research work presented in [10], involved post-graduate student performance considered dissimilar environments for a MCQ-oriented test. The tests were administered in a supervised in-class written manner or an unsupervised online manner. The results of their study show that, unsupervised online MCQ tests can be a feasible tool for evaluating post-graduate students if they meet best practice principles for online assessments.

In a pilot study [11], authors tried to measure university student insights of graded frequent assessments by means of a new questionnaire. Instead of a single final examination, several smaller assessment tasks are organized, each affecting the final grade. It is claimed that communication with students concerning the advantages of repeated assessments could alleviate some negative effects such as poor self-confidence or stress.

Although different assessment methods have been scrutinized frequently in recent years, more studies are required to yield a better awareness of the numerous methods in which assessment practice affects student learning process. Besides, most of these studies suffer from methodological flaws such as small sample size, lack of demographic comparisons, and significant differences in course materials and assessments.

To address these deficiencies, the objective of our study is to compare student learning outcomes of an in-class web-facilitated class with an asynchronous online class of a core networking undergraduate course considering diverse assessment methods. We also factored in demographics data using matching resources and assessments of both sections. Moreover, student's exposure to preferred assessments is an important source of information on the nature of the association between assessment and learning. The purpose of this study is to understand the effectiveness of assessments by discovering students' experiences with several dissimilar assessments in multiple sections of the core course.

### 3. Research Study

In this empirical study, we have collected data from four sections (01, 02, 04, and DL) of one course (IT 341) offered in Fall 2016 semester. One of these sections is an online class where most of the activities such as lecture slides, prerecorded lectures, home and lab assignments, etc. are done online via BB. However, students in this class are required to meet face-to-face (F2F) in a classroom on campus to take their midterm and final exams. In the online section, students communicate with their classmates, course instructor using blogs, and discussion forums. The other three F2F sections are all offered in-class, but uses BB to share course materials with students in similar manner as the DL section. Two of the four sections are taught by the same instructor, and the remaining sections are taught by different instructors.

Sessions of each section are conducted as follows: half of each session is led by an instructor dedicated to lecture materials (theoretical concepts), and the second half is devoted to completing skill-based lab assignments (working with Cisco Packet Tracer network simulator) for that week, which is led by the instructor and the Graduate Teaching Assistant (GTA). There is another assessment other than the classic home assignment, lab assignment and midterm/final exams. This assessment is called skill-based assessment (SA), which thoroughly evaluates students' hands-on skills as they work with the network simulator.

All sections have the same home assignment, lab assignments and SA. Also, lab sessions are conducted in the second half of each class, except section 01 in which the lab is conducted during the first half of the class. Although the concepts were the same for all the exams, we deliberately changed all sections exam structure to determine whether any pattern exists toward the students' performance. Since sections 01 and 04 are taught by the same instructor, so we considered the same exam structure of 25 Multiple Choice Questions (MCQs), fill-in-the-blank questions and a difficult numerical problem. Section 02 had 25 MCQs followed by two essay questions, and the DL section had 45 MCQs followed by a simple numerical problem. It is worth noting that, dissimilar essay questions and numerical problem questions were selected for the different sections. Students in all sections were given equal amount of time to complete their assessments, and were assessed by their respective instructor using the same grading rubric across all sections. The course grade was used to gauge students' learning performance.

A questionnaire was also developed to collect student demographic data. This includes data regarding gender, work status, age, commute distance to campus, networking experience, and marital status. Additionally, students were inquired about the type of question (MCQ or essay) they prefer for their test, and whether they like in-class or take home exams. Based on the demographics data, we addressed the following research questions in our analysis:

a) How does differing assessment structure affect student performance in different sections?
b) Does the modification of teaching method influence students' grades?
c) Is there a link between student's preference of a specific assessment method and their learning performance?
d) Is there an association between students' demographics and their performance?
e) How did the students perform in distance learning and in-class sections? Is there a noteworthy difference in their grades?

## 4. Analysis and Results

**A. Statistical Analysis (Welch t-test and ANOVA F test)**

We have compared all sections factoring in several assessment methods such as lab and assignments, skill-based assessment (SA) and midterm exam to determine which section is the best. In our analysis, we have used Welch t-test [12], which is a two-sample location test used to examine the hypothesis of any given two sections that have statistically equal means. Welch t-test is more appropriate when we have differing variances and unequal sample sizes.

The results of our comparison from Welch t-test are summarized in Table 1 where three p-values are greater than our assumed threshold value of 0.05, which means the performance of those sections are statistically equal to each other. Since most p-values are close to the threshold value, we have used ANOVA F test [13] to attain more accurate results across all the sections. We are using ANOVA model to compare the mean of four sections with each other using a parametric method (assuming sections follow Gaussian distribution).

Table 1: Results of Welch t-test Comparisons

| Comparison | p-value | Result |
|---|---|---|
| Section 01 & Section 02 | 0.0002838 | Significant difference |
| Section 01 & Section DL | 0.07663 | No significant difference |
| Section 02 & Section DL | 0.04873 | Significant difference |
| Section 01 & Section 04 | 0.4518 | No significant difference |
| Section 02 & Section 04 | 0.01022 | Significant difference |
| Section 04 & Section DL | 0.408 | No significant difference |

Assumptions regarding ANOVA F test are as follows:

- Independence of sections (given in our case).
- Normality – Residuals of the model are normally distributed.
- Homogeneity of variances.

Since we need to use ANOVA F test to compare the sections, we have to test our data based on the above assumptions. Therefore, Shapiro-Wilk [12] test was used to assess the normality assumption of residuals for the model. Our results show that the residuals are not normally distributed because p-value is less than 0.05. However, residuals are closer to Normal distribution, which is illustrated in Figure 1.

Bartlett test [13] was used to check the homogeneity of variances across all four sections. The p-value for Bartlett test is greater than 0.05, which proves the homogeneity of variances. Since we have tested all the assumptions, we were able to proceed with the ANOVA F test.

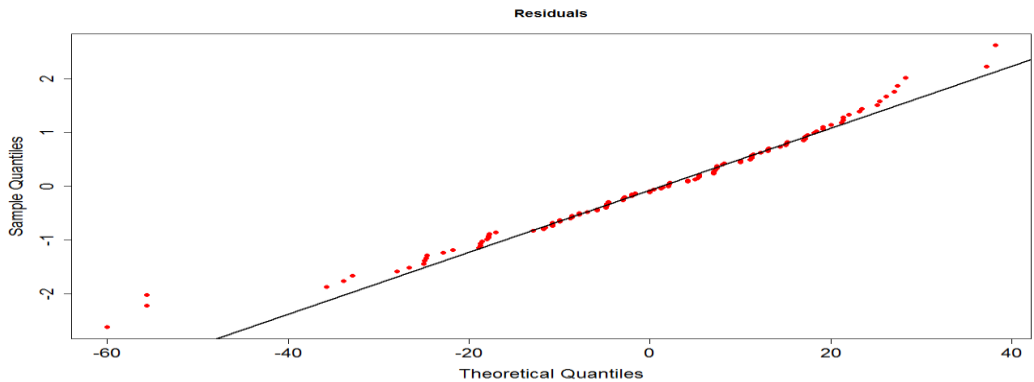

**Fig 1: Residual Normality Test Results**

Most types of nonnormality observed in the residual plot of Figure 1 and Shapiro-Wilk test result in a large ANOVA F test p-value as opposed to a smaller value. Typically, significant p-value (less than the threshold value) of ANOVA test can be taken as an evidence that not all of the means are equal. Since our ANOVA F test p-value is *0.002863*, we can accept the null hypothesis. Hence, there is statistically significant evidence that not all means are equal.

### B. Numerical Results and Findings

In order to analyze the data, we plotted the average of each section across different assessment methods as illustrated in Figure 2. The DL section did not perform well in labs as compared with the in-class sections perhaps because the students in F2F sections get in-class help, whereas the DL section students need to remotely contact the instructor and GTA for help/support.

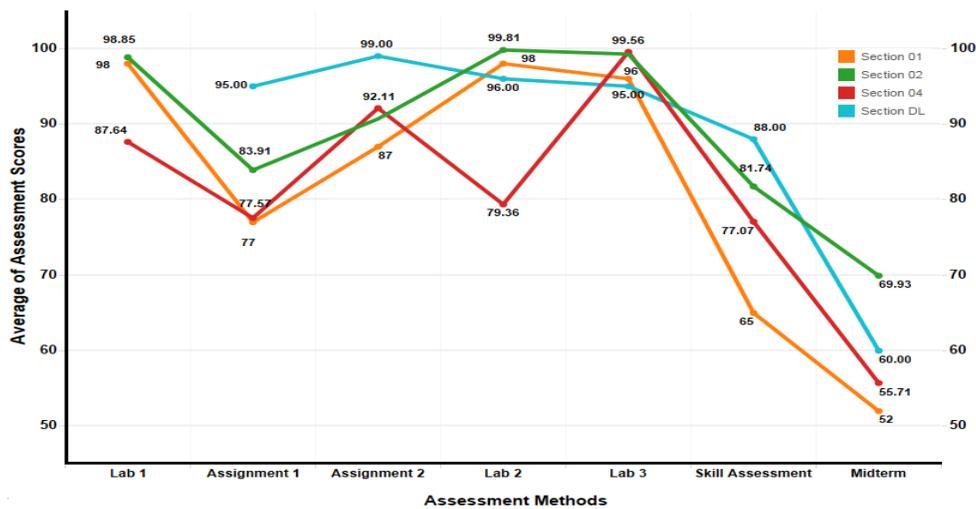

**Fig 2: Scores vs. Methods**

We assume the DL section performed best in SA because students took the exam from home and it is less stressful compared to in-class SA exam.

Figure 3 ascertains similar results from our demographic analysis based on the question, "Whether they preferred take home exams over in-class exams?"

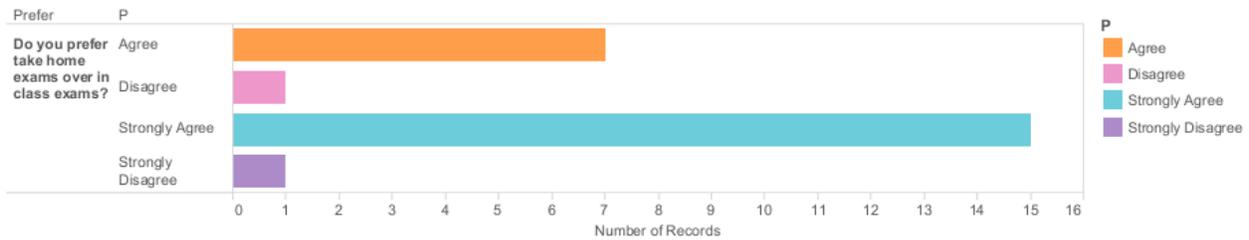

**Fig 3: Demographics of Exam Location Preference**

The same instructor teaches section 01 and section 04. However, section 01 performed worse in SA than section 04 maybe because unlike section 04 all the labs are conducted before the lectures in section 01. Although the SA covered the concepts and knowledge of the Labs, students struggled in SA because it did not have detailed instructions and the assessment was timed, whereas the labs provided step-by-step instructions and students worked on the labs in a much relaxed environment with a 1-week submission deadline.

We discovered that the students were more comfortable in expressing their knowledge in subjective questions and did best in section 02. We were also surprised to learn that most of the students favored MCQs as per the survey results presented in Figure 4 for section 02, but performed poorly in sections with the MCQs than the subjective questions.

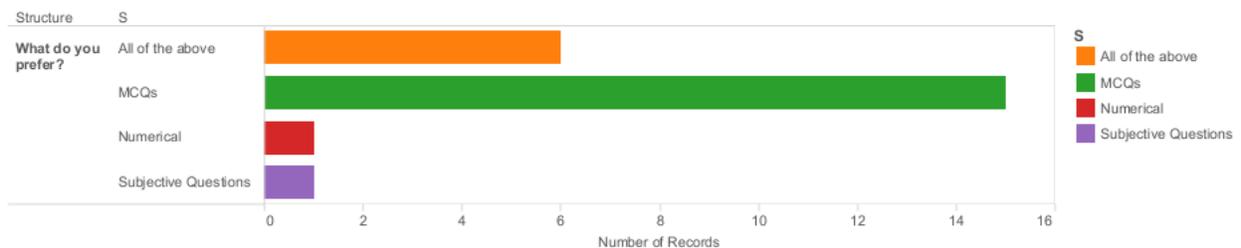

**Fig 4: Demographics of Exam Style Preference**

## 5. Conclusion and Future Works

In this study, we collected ample data from the learning management system (LMS), Blackboard (BB) Learn, in order to investigate the impact of several assessment outcomes for 4 sections of one course (IT 341). We also tried to understand the influence of demographics on student performance via a survey questionnaire. The numerical results reveal interesting insights regarding student achievement in the online and in-class sections. According to our analysis, students in DL section didn't do well enough as compared to the in-class sections due to lack of on-site supervisory help from their GTAs and/or instructors. Regarding skill-based assessment, it appears teaching method played an important role on the results of section 01, where lab sessions were conducted before the lecture. Finally, considering the fact that section

02 midterm test results outperform other sections, we reached the conclusion that students' performance increases with essay questions. For our future work, we will consider the impact of adjusted student BB access behavior, along with the significance of entirely modified second skill-based assessment and final exam, to be administered at the end of Fall 2016 semester.